\documentclass{iau} 
\usepackage{graphicx}
\usepackage{float}
\usepackage{epsfig}

\title[AGB stars mass--loss] {Mass--loss rates of cool evolved stars in M\,33 galaxy}

\author[Atefeh Javadi \& Jacco Th.\ van Loon]   
{Atefeh Javadi$^1$
 \and Jacco Th.\ van Loon$^2$}

\affiliation{$^1$School
of Astronomy, Institute for Research in Fundamental Sciences (IPM), P.\ O.\ Box 19395-5531, Tehran, Iran 
\\ email: {\tt atefeh@ipm.ir} \\[\affilskip]
$^2$Lennard-Jones Laboratories, Keele University, ST5 5BG, UK \\email: {\tt j.t.van.loon@keele.ac.uk }}

\pubyear{2022}
\volume{366} 
\jname{The origin of outflows in evolved stars}
\editors{L. Decin, A. Zijlstra,  \& C. Gielen}
\begin{document}

\maketitle

\begin{abstract}
We have conducted a near-infrared monitoring campaign at the UK InfraRed Telescope (UKIRT),
of the Local Group spiral galaxy M\,33 (Triangulum). In this  paper,
we present the dust and gas mass-loss rates by the pulsating Asymptotic Giant Branch (AGB)
stars and red supergiants (RSGs) across the stellar disc of M\,33. 

\end{abstract}

\firstsection 
\section{Introduction}

On the AGB, more than half of the mass is lost to the interstellar medium (ISM) in the form of a
dusty wind (van Loon et al.\ 2005). Mass loss is of great importance for stellar evolution and
the end products including supernovae, but also for the chemical enrichment of a galaxy. AGB stars
are the principal contributors of molecules and dust, and a major source of carbon and nitrogen. The
{\it Spitzer} mid--IR data allow us to derive accurate mass--loss rates. The luminosities and amplitudes
will then provide a relation between the mass--loss rate and the mechanical energy involved in the
pulsation (van Loon et al.\ 2006). Mass loss affects the pulsation period, which also depends on the
mantle mass. The amount of mass that has already been lost can thus be estimated from the period
and luminosity (Wood 2000). A statistical inventory of the mass loss along the AGB in different
metallicity range will yield the duration and strength of the mass loss, and thus provide feedback
intensities and timescales for chemical evolution models. The low--mass stars lose most of their
mass through dusty stellar winds, but even super--AGB stars and red supergiants lose $\sim40$\%
of their mass via a  stellar wind (Javadi et al.\ 2013). Furthermore, while more massive stars
(with birth masses $\gtrsim 8$ M$_\odot$)
are incapable of avoiding core collapse, mass loss
during the red supergiant (RSG) phase can severely deplete the mantle of the
star and even force a return to the blue
(Georgy 2012; Georgy et al.\ 2012). Fortunately, AGB stars and RSGs are relatively easy to detect, as they become
not only very luminous ($\sim10^{3.5-5.5}$ L$_\odot$) but also very red,
 and thus stand out at infrared (IR) wavelengths above other types of
stars within galaxies (Davidge 2000, 2018).

In this project, we aim to understand how galaxies such as our own have evolved to look the
way they do today. Our position within its dusty disc precludes such study in the
Milky Way, hence we turn to nearby spiral galaxy M\,33. We exploit the cool
variable stars that trace the endpoints of stellar evolution and are major sources of
dust. We monitored M\,33 with the UK InfraRed Telescope. Following our work on
the nucleus (Javadi et al.\ 2011), we will now [1] perform a census of cool variable stars across the
disc of M\,33; [2] reconstruct the star formation history across M\,33 (and other
nearby galaxies) and [3] quantify the return of matter throughout M\,33. 

\section{Why M\,33  galaxy?}

M\,33 is the nearest spiral galaxy besides the Andromeda galaxy, and seen under a
more favourable angle. This makes M\,33 ideal to study the structure and evolution
of a spiral galaxy. We will thus learn how our own galaxy the Milky Way formed
and evolved, which is difficult to do directly due to our position within its dusty
disc. 

The methodology consists of three different phases: [1] firstly, we identify long period variables stars (LPVs)
(Javadi et al.\ 2015); [2] secondly, we uniquely relate their brightness to their birth mass, and use the birth mass
distribution to reconstruct the star formation history (SFH) (Javadi et al.\ 2017); [3] thirdly, we measure the excess
infrared emission from dust produced by these stars, to  estimate the amount of
matter they return to the interstellar medium in M\,33 (Javadi et al.\ 2013).

\section{The data we use}

To derive the mass--loss rates of evolved stars we make use
of two data sets; our own near--IR data in the J, H and
$K_s$ bands (Javadi et al.\ 2015) and archival mid--IR {\it Spitzer}
data at 3.6, 4.5 and 8 $\mu$ m (McQuinn et al.\ 2007).

\subsection{Near--IR data}
The project exploits our large observational campaign between 2003-2007, over
100 hr on the UK InfraRed Telescope.  The observations were done  in
the $K_s$--band ($\lambda$= 2.2 $\mu$m) with occasionally observations in
J-- and H--bands ($\lambda$= 1.28 and 1.68 $\mu$m, respectively) for the
purpose of obtaining colour information. The photometric catalogue comprises 403\,734 stars,
among which 4643 stars were
identified as LPVs -- AGB stars, super--AGB stars and RSGs. 

\subsection{Mid--IR data}
Using five epochs of {\it Spitzer} Space Telescope imagery in the 3.6--, 4.5-- and
8 $\mu$m bands, variables have been identified by McQuinn et al.\ (2007), 
using a similar method to that we used ourselves.

Of the stars in common, 985 stars were identified as variables in both surveys,
but two were saturated and therefore excluded from further
analysis. This means that 3658 of the WFCAM variable stars were not identified
as variables in the {\it Spitzer} survey, which is mainly because of the
limitation of {\it Spitzer} in detecting the fainter, less dusty variable red
giants. On the other hand, the {\it Spitzer} survey identified 2923 variables,
suggesting a one--third completeness level of the WFCAM variable star survey --
this agrees with our internal assessment from a comparison between the WFCAM
and UIST data on the central square kpc (Javadi et al.\ 2015). Generally, both surveys do
well in detecting dusty variable AGB stars (and RSGs); this is crucial to
estimate mass--loss rates based on IR photometric data.

\section{From LPVs luminosities to the star formation history}
In the final stage of stellar evolution, low-- and intermediate mass (0.8--8  M$_\odot$) 
stars enter the AGB phase (Marigo et al.\ 2017) and high mass (M$\gtrsim$8 M$_\odot$) stars enter the 
RSG phase (Levesque 2010). These two phases of stellar evolution 
are characterized by strong radial pulsation of cool atmosphere layers, making them  identifiable as LPVs in the 
photometric light curves (Ita 2004; Yuan et al.\ 2018; Goldman et al.\ 2019). The LPVs 
({AGB--stars, super--AGB stars and RSGs}) are at 
the end--points of their evolution, and their luminosities directly reflect their birth
mass (Javadi et al.\ 2011). Stellar evolution models provide this relation. The distribution of LPVs
over luminosity can thus be translated into the star formation history, assuming a standard initial
mass function. Because LPVs were formed as recently as $<$ 10 Myr ago and as long ago as $>$ 10 Gyr,
they probe almost all of cosmic star formation. We have successfully used this new technique in M\,33
(Javadi et al. 2011, 2017) using the Padova models which also provide the lifetimes of the
LPV phase (Marigo et al.\ 2017).

\section{Modelling the spectral energy distribution}

\begin{figure}[b]
\begin{center}
 \includegraphics[width=5.1in]{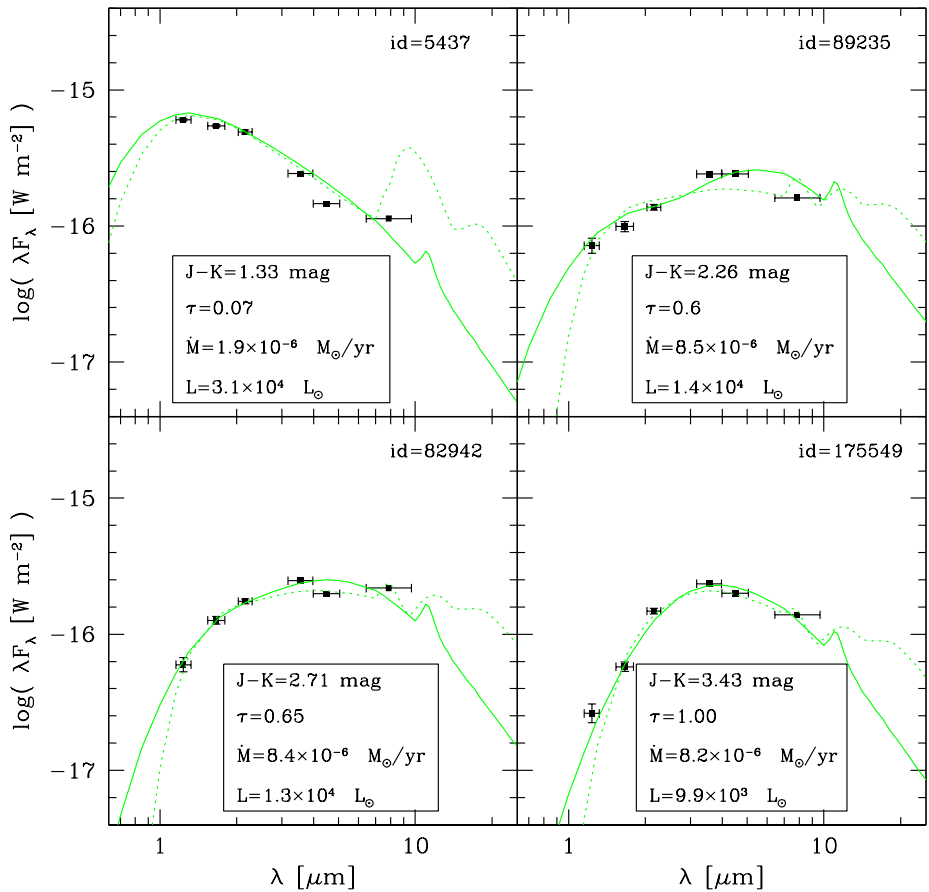} 
\caption{Example SEDs of presumed carbon stars. The horizontal "errorbars"
on the data represent the width of the photometric passbands. The best
matching SEDs modelled with {\sc dusty} are shown with solid lines. For
comparison, the best matching fits using silicates are shown with dotted
lines.}
   \label{fig1}
\end{center}
\end{figure}

\begin{figure}[b]
\begin{center}
 \includegraphics[width=5.1in]{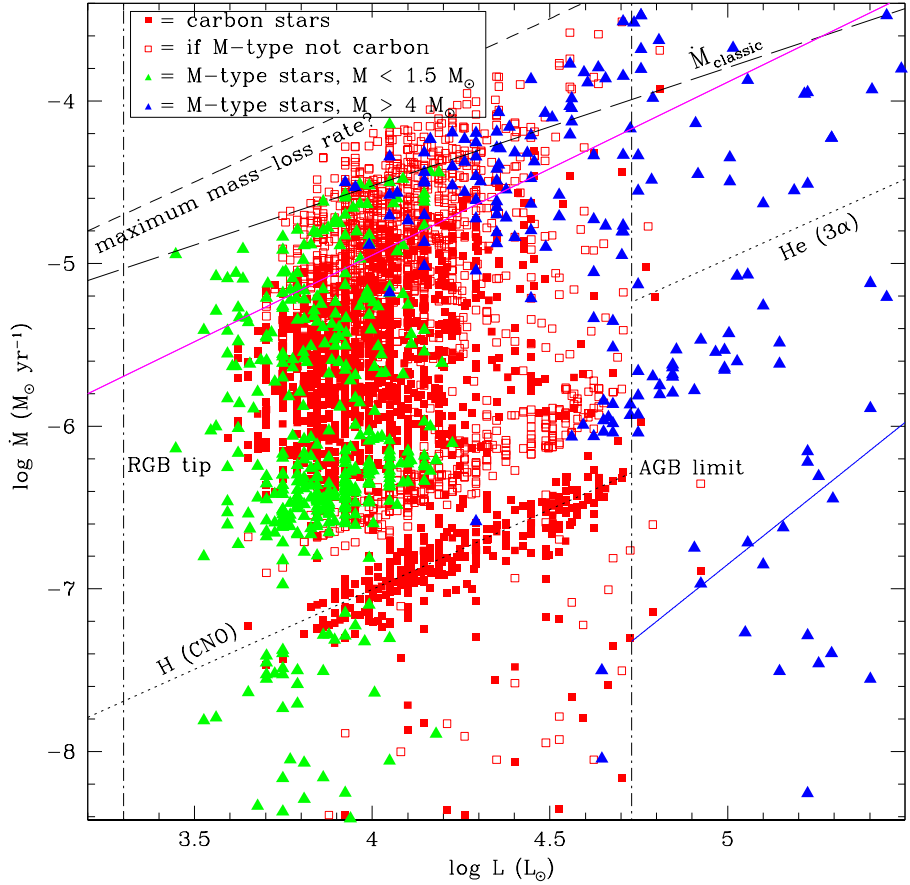} 
\caption {Mass--loss rate vs.\ luminosity, modelled with {\sc dusty} for
low--mass AGB stars (green triangles), intermediate--mass carbon stars (red
squares) and massive AGB stars and RSGs (blue triangles). The open red squares
show the results if the presumed carbon stars are presumed to be oxygen--rich
instead. The vertical dash--dotted lines mark the tip luminosity of the first ascent
red giant branch (RGB) and the classical limit of the most massive AGB stars
(excluding the effects of Hot Bottom Burning). The dotted lines trace the
mass--consumption rates by shell hydrogen burning (CNO cycle) on the AGB and
core helium burning (triple--$\alpha$ reaction) in RSGs. The dashed lines trace
the limits to the mass--loss rate in dust-driven winds due to single scattering
(classic) and multiple scattering (maximum?; see van Loon et al.\ 1999). The magenta line
traces the fit to the mass--loss rate vs.\ luminosity presented in Goldman et
al.\ (2017), whilst the blue line traces the relation found by Verhoelst et
al.\ (2009) for Galactic RSGs.}
   \label{fig2}
\end{center}
\end{figure}

Spectral energy distributions (SEDs) contain information about the stellar luminosity, 
temperature, metal content, surface gravity
and extinction. If sampled over a sufficient range in wavelength, 
employing accurate stellar spectral
templates allows to retrieve some or all of these parameters.
To model SEDs of WFCAM variables we used the publicly available dust radiative
transfer code {\sc DUSTY} (Ivezi\'c \& Elitzur 1997). All variables with at
least two measurements in near--IR bands (K$_{\rm s}$ and J and/ or H) and two
mid--IR bands (3.6, 4.5 and/or 8 $\mu$m) were modelled ($\sim$ 2000 stars). 
{\sc DUSTY} calculates the radiation transport in a dusty envelope. We fixed
the input temperatures of the star and of the dust at the inner edge of the
circumstellar envelope, at 3000 and 900 K, respectively. The density structure
is assumed to follow from the analytical approximation for radiatively driven
winds (Ivezi\'c, Nenkova \& Elitzur 1999). This obviates the need to assume or
measure the outflow velocity, as it is implicit in the relation between
luminosity, optical depth, gas--to--dust mass ratio and mass--loss rate.
We used amorphous carbon dust (Hanner 1988) and a small amount
(15 per cent) of silicon carbide (P\'egouri\'e 1988) for carbon stars, and astronomical
silicates (Draine \& Lee 1988) for M--type stars (Fig.\ 1). 
Because a sub--set of AGB stars, carbon stars have a different type of
circumstellar dust, we must try to identify which stars are likely to be
carbon stars. In the absence of spectroscopic confirmation for most of these,
and the limited constraints we have from photometry, we resort to making use
of theoretical expectations. Correcting the observed colours for the effect of
circumstellar dust, we obtain an intrinsic K--band brightness. Using stellar
evolution models (Marigo et al.\ 2017) we convert this into a birth
mass, given that these are highly evolved stars that will not evolve much in
luminosity. The mass range for AGB stars to
become carbon stars spans $\sim1.5$--4 M$_\odot$.

\subsection{Mass--loss rate}
For our complete sample (Javadi et al.\ 2015),
some dependence of mass--loss rate on luminosity is seen (Fig.\ 2); the maximum
mass--loss rate increases with luminosity and the highest mass--loss rates are
generally achieved by the most luminous, most massive large--amplitude variable
stars. This confirms earlier studies in the central region of M\,33 (Javadi et al.\ 2013) 
and in the Magellanic Clouds (Srinivasan et al.\
2009). The mass--loss rates for M--type AGB stars and RSGs are similar to those
found in the Solar Neighbourhood (a few $\times10^{-5}$ and $10^{-7}$--$10^{-4}$
M$_\odot$ yr$^{-1}$, respectively; Jura \& Kleinmann 1989). The mass--loss rates
for presumed carbon stars are also in good agreement with those found in the
Milky Way (a few $\times10^{-5}$ M$_\odot$ yr$^{-1}$; Whitelock et al.\ 2006)
and in the Magellanic Clouds ($\sim10^{-5}$ M$_\odot$ yr$^{-1}$; Gullieuszik et
al.\ 2012).

It is reassuring to see that the RSGs
 (certainly stars well above the AGB limit of $\log L/L_\odot=4.73$ --
Wood, Bessell \& Fox (1983))
are generally oxygenous; that the least luminous stars are too, and that the
maximum mass--loss rate increases with luminosity (in fact rather steeply).
Oxygenous stars around -- or slightly fainter than -- the AGB limit with very
high mass--loss rates are probably massive AGB stars, the equivalent of (most
of) the OH/IR stars that are found in the LMC (Marshall et al.\ 2004).

\section{On going works and conclusion remarks}
Comparison  of the total mass return rate from dusty evolved stars across the galactic 
disc of M\,33 ($\approx $0.1 M$_\odot $yr$^{-1}$; Fig.\ 2) and recent star formation rate 
( $\xi=0.45\pm0.10$ M$_\odot $yr$^{-1}$; Javadi et al.\ 2017), 
suggests that for star formation to continue beyond the next Gyr or
so, gas must flow into the disc of M 33, via cooling flows
from the circum--galactic medium and/or by inward migration from gas reservoirs
in the outskirts of the disc (Javadi et al.\ in prep).

In order to gain a comprehensive understanding of galaxy formation and evolution in the Local Group, 
 recently we have conducted an optical monitoring survey of the majority of nearby dwarf galaxies 
 with Isaac Newton Telescope (INT) to identify LPVs (Saremi et al.\ 2019, 2020, 2021; Navabi el al.\ 2021). 
 This research is very important from both theoretical and observational perspectives: First,
 it will give an unprecedented map of the temperature and radius variations as a function 
 of luminosity and metallicity for mass-losing stars at the end of their evolution, which
 places important constraints on stellar evolution models and which is a vital ingredient 
 in the much sought-after description of the mass-loss process. Second, from observational
 prospective, this research will gather independent diagnostics of the SFHs of different
 types of dwarf galaxies found in different environments, which help build a detailed
 picture of galaxy evolution in the nearby galaxies.


\begin{thebibliography}{}
\bibitem[Davidge (2000)]{Davidge00}
Davidge T., 2000, AJ, 119, 748
\bibitem[Davidge (2018)]{Davidge18}
Davidge T., 2018, ApJ, 856, 129
\bibitem[Draine \& Lee (1984)]{Draine84}
Draine B.\ T., Lee H.\ M., 1984, ApJ, 285, 89
\bibitem[Georgy (2012)]{Georgy12}
Georgy C., 2012, A\&A, 538, L8
\bibitem[Georgy et al.(2012)]{Georgy12x}
Georgy C., Ekstr\"om S., Meynet G., Massey P., Levesque E.\ M., Hirschi R.,
Eggenberger P., Maeder A., 2012, A\&A, 542, A29
\bibitem[Goldman et al.(2017)]{Goldman17}
Goldman S.\ R.\ et al., 2017, MNRAS, 465, 403
\bibitem[Goldman et al.\ (2019)]{Goldman19}
Goldman S.\ R., 2019, ApJ, 877, 49
\bibitem[Gullieuszik et al.(2012)]{Gullieuszik12}
Gullieuszik M.\ et al., 2012, A\&A, 537A, 105
\bibitem[Hanner (1988)]{Hanner88}
Hanner M.\ S., 1988, NASA Conf.\ Pub.\ 3004, 22
\bibitem[Ita et al.\ (2004)]{Ita04}
Ita Y., et al.\, 2004, MNRAS, 353, 705
\bibitem[Ivezi\'c \& Elitzur (1997)]{Ivezic97}
Ivezi\'c \v{Z}, Elitzur M., 1997, MNRAS, 287, 799
\bibitem[Ivezi\'c, Nenkova \& Elitzur (1999)]{Ivezic99}
Ivezi\'c \v{Z}, Nenkova M., Elitzur M., 1999, {\sc dusty} User Manual
(University of Kentucky)
\bibitem[Javadi et al.\ (2011)]{Javadi11}
Javadi A., van Loon J.\ Th., Mirtorabi M.\ T., 2011, MNRAS, 414, 3394
\bibitem[Javadi et al.\ (2013)]{Javadi13}
Javadi A., van Loon J.\ Th., Khosroshahi H.\ G., Mirtorabi M.\ T., 2013, MNRAS, 432, 2824
\bibitem[Javadi et al.\ (2015)]{Javadi15}
Javadi A., Saberi M., van Loon J.\ Th., Khosroshahi H.\ G., Golabatooni N., Mirtorabi M.\ T., 2015, MNRAS, 447, 3973
\bibitem[Javadi et al.\ (2017)]{Javadi17}
Javadi A., van Loon J.\ Th., Khosroshahi H.\ G., Tabatabaei F., Hamedani Golshan R., Rashidi M., 2017, MNRAS, 464, 2103
\bibitem[Jura \& Kleinmann (1989)]{Jura89}
Jura M., Kleinmann S.\ G., 1989, ApJ, 341, 359
\bibitem[Levesque et al.\ (2005)]{Levesque05}
Levesque E. M., Massey P., Olsen K. A. G., Plez B., Josselin E., Maeder A., Meynet G., 2005, ApJ, 628, 973
\bibitem[Marigo et al.\ (2017)]{Marigo17}
Marigo P.\ et al., 2017, ApJ, 835, 19
\bibitem[Marshall et al.(2004)]{Marshall04}
Marshall J.\ R., van Loon J.\ Th., Matsuura M., Wood P.\ R., Zijlstra A.\ A.,
Whitelock P.\ A., 2004, MNRAS, 355, 1348
\bibitem[McQuinn et al.(2007)]{Mcquinn07}
McQuinn K.\ B.\ W.\ et al., 2007, ApJ, 664, 850
\bibitem[Navabi et al.\ (2021)]{Navabi21}
Navabi M., Saremi E., Javadi A., Noori M., van Loon J.\ Th., Khosroshahi H.\ G., McDonald I., Alizadeh M., Danesh A., Gozaliasl G., Molaeinezhad A., Parto T., Raouf M., 2021, ApJ, 910, 127
\bibitem[P\'egouri\'e (1988)]{Pegourie88}
P\'egouri\'e B., 1988, A\&A, 194, 335
\bibitem[Saremi et al.\ (2019)]{Saremi19}
Saremi E., Javadi A., van Loon J.\ Th., Khosroshahi H.\ G., Rezaeikh S., Hamedani Golshan R., Hashemi S.\ A., 2019, Proceedings of IAU Symposium, 344, 125
\bibitem[Saremi et al.\ (2020)]{Saremi20}
Saremi E.\ et al., 2020, ApJ, 894, 135 
\bibitem[Saremi et al.\ (2021)]{Saremi21}
Saremi E, Javadi A., Navabi M., van Loon J.\ Th., Khosroshahi H.\ G., Bojnordi Arbab B., McDonald I., 2021, ApJ, 923, 164
\bibitem[Srinivasan et al.(2009)]{Srinivasan09}
Srinivasan S.\ et al., 2009, AJ, 137, 4810
\bibitem[van Loon et al.(1999)]{vanLoon99}
van Loon J.\ Th., Groenewegen M.\ A.\ T., de Koter A., Trams N.\
R., Waters L.\ B.\ F.\ M., Zijlstra A.\ A., Whitelock P.\ A., Loup
C., 1999, A\&A, 351, 559
\bibitem[van Loon et al.(2005)]{vanLoon05a}
van Loon J.\ Th., Cioni M.-R.\ L., Zijlstra A.\ A., Loup C., 2005, A\&A, 438,
273
\bibitem[van Loon et al.(2006)]{vanLoon06}
van Loon J.\ Th., Marshall J.\ R., Cohen M., Matsuura M., Wood P.\ R.,
Yamamura I., Zijlstra A.\ A., 2006, A\&A, 447, 971
\bibitem[Verhoelst et al.(2009)]{Verhoelst09}
Verhoelst T., Van der Zypen N., Hony S., Decin L., Cami J., \& Eriksson K.,
2009, A\&A, 498, 127
\bibitem[Whitelock et al.(2006)]{Whitelock06}
Whitelock P.\ A., Feast M.\ W., Marang F., Groenewegen M.\ A.\ T., 2006,
MNRAS, 369, 751
\bibitem[Wood, Bessell \& Fox (1983)]{Wood86}
Wood P.\ R., Bessell M.\ S., Fox M.\ W., 1983, ApJ, 272, 99
\bibitem[Wood (2000)]{Wood00}
Wood P.\ R., 2000, PASA, 17, 18
\bibitem[Yuan (2018)]{Yuan18}
Yuan W., Macri L.\ M., Javadi A., Lin Z., Huang J.\ Z., 2018, AJ, 156, 112
\end{thebibliography}
\end{document}